\documentclass[12pt]{article}
\usepackage{epsf}
\usepackage{graphicx}
\usepackage{overcite}

\usepackage[dvips]{epsfig,geometry}

\begin{document}

\title{Continuum model for polymers with finite thickness}

\author{D. Marenduzzo $^{\dag,\ddag}$, C. Micheletti$\S$, H.
Seyed-allaei$\S$, A. Trovato$\|$ and A. Maritan$\|$}

\maketitle

\noindent{$\dag$ Department of Physics, Theoretical Physics,
University of Oxford, 1 Keble Road, Oxford OX1 3NP, England\\
$\ddag$ Mathematics Institute, University of Warwick, CV4 7AL, England\\
$\S$ International School for Advanced Studies (SISSA) and INFM,
Via Beirut 2-4, 34014 Trieste, Italy\\
$\|$ INFM and Dip. di Fisica `G. Galilei', Universit\'a di
Padova, Via Marzolo 8, 35100 Padova, Italy}




\begin{abstract}
We consider the continuum limit of a recently-introduced model for
discretized thick polymers, or tubes. We address both analytically and
numerically how the polymer thickness influences the decay of
tangent-tangent correlations and find how the persistence length
scales with the thickness and the torsional rigidity of the tube
centerline. At variance with the worm-like chain model,
the phase diagram that we obtain for a continuous tube is richer;
in particular, for a given polymer thickness there exists a threshold value
for the centerline torsional rigidity separating a simple exponential
decay of the tangent-tangent correlation from an oscillatory one.
\end{abstract}


Experimental studies of biopolymers have constantly stimulated the
search for schematic models apt for reproducing the observed kinetic
and thermodynamic behaviour. In recent years two types of biopolymers
have attracted most of these efforts: DNA and proteins. The interest
in the former has been sparked by the introduction of single-molecule
experiments which probed the elastic response of DNA upon stretching
\cite{bustamante}. Considerable progress in the rationalization of
these experiments was made thanks to the worm-like-chain (WLC) model
\cite{Marko1,Calu} where the polymer is described as a continuous
centerline possessing an effective bending \cite{Marko1} and/or
twisting \cite{Moroz1,Mezard1} rigidity. For protein modelling,
instead, one of the goals is to capture the main physico-chemical
forces responsible for driving the folding process towards the native
state. Typical coarse-grained models adopt sophisticated energy
functionals (often with hundreds of parameters) in order to reproduce
the observed variety of protein folds. It has been recently argued,
however, that the overburdening of the energy function can be avoided
by modelling explicitely the intrinsic thickness of
proteins\cite{Nature,Proteins,CPU,PNAS}. As we discuss later this is
achieved through the introduction of suitable three-body interactions
among triplets of points constituting the polymer centerline
\cite{buck1,maddocks,JSP,CPU,RMP}.  It is physically appealing that
the thick-chain model has proved valuable also for the case of DNA in
applications ranging from the characterization of knotted DNA
molecules \cite{katrich} to the thermodynamics of DNA packaging inside
viral capsids and DNA elastic response\cite{capsid}.

By necessity, all numerical implementations of these models rely in
the discretization of the polymer centerline into a succession of
beads whose ``natural'' spacing is typically suggested by the
intrinsic granularity of the polymer itself (e.g. the separation of
consecutive $C_\alpha$'s for proteins or the base-pair spacing in
dsDNA). From a theoretical perspective it is therefore desirable to
characterize the thick chain model in the continuum limit, where the
bead spacing tends to zero (analogously to the WLC limit of the
Kratky-Porod model). This continuum limit has, so far, been considered
only for characterizing the limited repertoire of ideal knots.
Motivated by the potentially-wide range of applicability of the
thick-polymer model, in this Letter we take the perspective of
addressing the statistical mechanics of generalized thick-polymer
models in the continuum limit. In particular, we introduce in the
Hamiltonian a penalty for the geometrical torsion of the tube
centerline and, initially, consider the constraints induced by the
finite polymer thickness only at a local level, a simplification
usually adopted to allow analytical progress
\cite{Moroz1,Mezard1}. From the exact analysis it emerges that
accounting for the centerline torsional rigidity term (1) allows to
get a finite persistence length in the limit of a continuous thick
polymer and (2) introduces a novel feature in the behavior of the
tangent-tangent correlation function, namely the presence of a
threshold value for the torsional rigidity which separates a monotonic
decay from an oscillatory one.  Finally, numerical Monte Carlo
simulations are employed to show that this behavior persists also when
the tube constraint is enforced at non local level. These findings
highlight the rich behaviour of models where the thickness is treated
explicitly. As a comparison we consider the case of a WLC in the
presence of penalty for the centerline geometrical torsion. It is
found that this model exhibits {\em either} a simple exponential decay
{\em or} an oscillatory one (with singular behaviour in the continuum
limit) depending, respectively, on the absence or presence of the
torsional rigidity but independently of its strength. In relation to
the behaviour observed here, it should also be noted that Panyukov and
Rabin\cite{Rabin1} have considered, in place of the whole equilibrium
ensemble, a rod-like chain fluctuating around a stress-free helical
conformation. Under these conditions they could observe a change from
an oscillatory to a simple exponential decay by increasing the
fluctuation strength.

\begin{figure}
\centerline{\includegraphics[width=3.0in,height=1.3in]{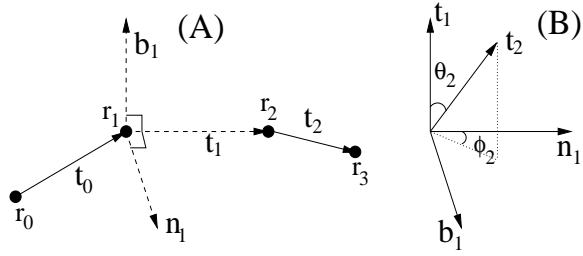}}
\caption{Frenet reference frame for a discrete bead model. Note that
$\theta_i \in [0,\pi]$ is defined for $1\le i\le N-2$, whereas
$\phi_i \in [0, 2\pi]$ is defined only for $2\le i\le N-2$.}
\label{fig:reframe}
\end{figure}

We model a polymer chain by means of a set of $N$ consecutive beads,
$\{\vec r_0 ... \vec r_{N-1}\}$ connected by bonds of fixed length,
$a$.  The succession of beads constitutes the centerline for our thick
polymer. We shall denote with $\Delta$ and $\kappa_t$ the thickness of
the chain and the torsional rigidity, respectively. Although we shall
first focus on the case $\kappa_t=0$ we will develop a formalism
general enough to be used also in the WLC with torsional penalty.  By
analogy with the Frenet reference frame for continuous curves
\cite{Kamien}, we define an orthonormal set associated to each bead,
formed by the local {\em tangent}, $\hat t_i\equiv (\vec r_{i+1}-\vec
r_i) / a$, {\em binormal}, $\hat b_i \equiv \hat t_i \wedge \hat
t_{i-1}/ |\hat t_i \wedge \hat t_{i-1} |$ and {\em normal}, $\hat n_i
\equiv \hat t_i \wedge \hat b_i$ (see Fig. \ref{fig:reframe}a).
It is possible to write recursion equations relating the reference
axes for bead $i$ in terms of those for bead $i-1$, using the polar
angles $\theta_{i}$ and $\phi_{i}$ as in Fig. \ref{fig:reframe}b:
\begin{eqnarray}
\left\{\begin{array}{l l}
&\hat b_{i} =
\cos{\phi_{i}}\hat b_{i-1} - \sin{\phi_{i}}\hat n_{i-1}\:,\\
&\hat t_{i} = \sin{\theta_{i}}\sin{\phi_{i}}\hat b_{i-1} +
\cos{\theta_{i}}\hat t_{i-1} +
\sin{\theta_{i}}\cos{\phi_{i}}\hat n_{i-1}\:,\\
&\hat n_{i} =  \cos{\theta_{i}}\sin{\phi_{i}}\hat b_{i-1} -
\sin{\theta_{i}}\hat t_{i-1} +
\cos{\theta_{i}}\cos{\phi_{i}}\hat n_{i-1}\:.
\end{array}
\right. \nonumber
\end{eqnarray}
\noindent Quite generally the joint probability distribution of angles,
${\mathcal{P}}(\theta_1, \theta_2, \phi_2, \theta_3, \phi_3, \dots )$,
resulting from the canonical average, will depend on the whole
ensemble of interactions including the steric ones. However,
for the simplified case where the effects of the polymer thickness are
treated only locally (as for twist and bending rigidity) then
$\mathcal{P}$ factorises in terms of the probability distributions
$\rho(\theta_i, \phi_i)$ for each pair of angles $\theta_i, \phi_i$:
\begin{equation}
{\mathcal{P}}(\theta_1, \phi_2,\theta_2,\phi_3,\theta_3, \dots ) =
\rho_{\theta}(\theta_1)\prod_{i=2}^{N-2}\rho(\theta_i, \phi_i) \: .
\label{prob}
\end{equation}
\noindent Since we are considering a uniformly-thick homopolymer, the same
probability distribution, $\rho\left(\theta,\phi\right)$, is involved
for all beads. In the following the averages weighted with $\rho$ will
be denoted as $\langle \cdot \rangle$, while those calculated with
respect to $\mathcal{P}$ will be written as $\langle \cdot
\rangle_{\mathcal{P}}$. The factorization leads to a straightforward
characterization of the decay along the chain of expectation values
such as $f_{i} \equiv \langle \hat b_{i}\cdot \vec x \rangle_{\mathcal{P}}$,
$g_{i} \equiv \langle \hat t_{i}\cdot \vec x \rangle_{\mathcal{P}}$,
$h_{i} \equiv \langle \hat n_{i}\cdot \vec x \rangle_{\mathcal{P}}$,
where $\vec{x}$ could be, e.g., $\hat t_1$, $\hat b_1$, $\hat n_1$. 
In fact, $f$, $g$ and $h$ at location $i+1$ are obtained from those at
site $i$ by the application of the following transfer matrix:
\begin{eqnarray}
\mathbf{T} = \left(\begin{array}{c c c}
\left\langle \cos{\phi}\right\rangle & 0 &  - \left\langle \sin{\phi}\right\rangle  \\ 
\langle \sin{\theta}\sin{\phi} \rangle  & \langle \cos{\theta} \rangle &  
\langle \sin{\theta}\cos{\phi} \rangle \\
\left\langle \cos{\theta\sin{\phi}}\right\rangle  &
-\left\langle \sin{\theta}\right\rangle & 
\left\langle \cos{\theta\cos{\phi}}\right\rangle
\end{array}
\right)\:.
\label{tm}
\end{eqnarray}
\noindent If the eigenvalues of $\mathbf{T}$ are real, the decay of $f$, $g$ and
$h$ will be monotonic, while if they are imaginary there will be an
oscillatory modulation.

We now consider two further simplifying assumptions: (i) the 'bond'
angle $\theta$ and the 'dihedral' angle $\phi$ contribute
independently to the probability distribution $\rho(\theta,\phi) =
\rho_{\theta}(\theta) \rho_{\phi}(\phi)$, and (ii) the system is
invariant for chirality flipping $\rho_{\phi}(\phi) =
\rho_{\phi}(-\phi)$.  In this case,
$\left\langle\sin{\phi}\right\rangle=0$, and the transfer matrix
$\mathbf{T}$ becomes block diagonal with an eigenvalue $\lambda_1 =
\left\langle\cos{\phi}\right\rangle$, so that $\left\langle
\hat{b}_n\cdot\hat{b}_1\right\rangle = \lambda_1^{n-1} =
\exp\left[-a\left(n-1\right)/\xi_b\right]$ decays exponentially with a
correlation length $\xi_b = -a/\ln\langle\cos\phi\rangle$. The
remaining two eigenvalues are the solutions of the second order
equation $\lambda^2 - b \lambda + c = 0$ with $b =
\langle\cos\theta\rangle \left[1+\langle\cos\phi\rangle\right]$, $c =
\langle\cos\phi\rangle \left[\langle\cos\theta\rangle^2 +
\langle\sin\theta\rangle^2\right]$.  The relevant quantity which
discriminates between different decay properties is $\Gamma \equiv b^2
- 4c$ .  If $\Gamma > 0$, the two solutions $\lambda_{2,3} =
(\pm\sqrt{\Gamma}+b)/2$ are real and the correlation function for
tangent/normal vectors decays exponentially to zero, with the
correlation length $\xi_t = -a/\ln\lambda_2$ being controlled by the
largest eigenvalue. If $\Gamma < 0$, the two solutions are complex
conjugate, and the tangent-tangent correlation function exhibits an
oscillatory decay:
\begin{equation}
\langle \hat t_{n+1} \cdot \hat t_1 \rangle = 
\frac{\cos\left[a/\chi_0 + a\,n/\chi\right]}{\cos\chi_0}e^{-a\, n/\xi_t} \; ,
\label{oscde}
\end{equation}
where $\xi_t = -2a/\ln c$, $\chi = a/\arctan\left(\sqrt{-\Gamma}/b\right)$,
and $\chi_0$ depends on initial conditions.

Within this general framework we now consider different specific
examples. We first focus on the case of a WLC subject to a penalty for
the geometrical torsion; the corresponding Hamiltonian is:
\begin{equation}
{\cal H}_{\rm WLC} =  \frac{\kappa_b}{2a} \sum_i | \hat t_{i+1} - \hat t_{i}|^2
+ \frac{\kappa_t}{2a} \sum_i | \hat b_{i+1} - \hat b_{i}|^2\:,
\label{RLC}
\end{equation}
\noindent where $\kappa_b$, $\kappa_t$, are the bending and torsional
rigidity, respectively, defined in such a way to get back the usual
Hamiltonian in the continuous limit, $a\rightarrow0$. We emphasize the
fact that the Hamiltonian (\ref{RLC}) is entirely specified by the
centerline geometry. The approach therefore differs in spirit from
those used to model the elasticity of rod-like chains. In these
contexts, starting from a stress-free refecence configuration, one
introduces a ``material reference frame'' which is used to keep track
not only of the deformations of a given reference frame, but also of
the twist around the centerline \cite{Kamien}. Although this latter
information is clearly not available in the model of eqn. (\ref{RLC}) it
is worth to considering energy-functions relying uniquely on the
knowledge of the centerline. In fact, for the important class of
biopolymers constituted by proteins, it is well known that the
knowledge of a protein's centerline (the $C_\alpha$ trace) and
sequence composition allows to reconstruct the whole protein structure
with high accuracy.

\begin{figure}[htbp]
\centerline{\includegraphics[width=2.4in]{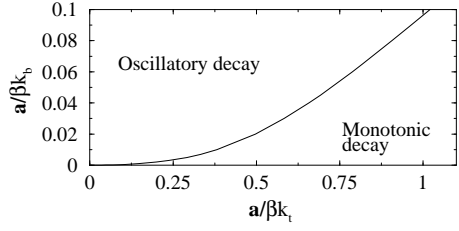}}
\caption{Boundary separating the oscillatory from the exponential
  decay of the tangent-tangent correlation. }
\label{fig:phdiag}
\end{figure}

For the case of Hamiltonian (\ref{RLC}) the probability distributions
for bond and dihedral angles taking, of course, into account the
inverse temperature $\beta$ is found to be: $\rho_{\theta}\left(\theta\right) =
\sin\theta \exp\left[ \beta \frac{\kappa_b}{a} \cos\theta \right]$;
$\rho_{\phi}\left(\phi\right) = \exp\left[ \beta \frac{\kappa_t}{a}
\cos\phi \right]$. All averages appearing in the transfer matrix
elements of eq. (\ref{tm}) can formally be expressed by means of
modified Bessel functions which, in turn allow to identify the
boundary, $\Gamma =0$, separating the oscillatory from the monotonic
decay of tangent correlations, see Fig. \ref{fig:phdiag}. In the
continuum limit, $a\rightarrow0$, the angles $\theta$ and $\phi$
contributing significantly to the average come from a region centred
around zero and of width $\sqrt{{a}/{\beta \kappa_b}}$ and
$\sqrt{{a}/{\beta \kappa_t}}$, respectively and the equation for the
boundary is $a/\beta \kappa_b \approx (a / \beta \kappa_t)^2 /8
\pi$. This implies that, for any finite value of the torsional
rigidity, in the continuum limit the tangent-tangent correlation
function always decays in an oscillatory way. The period of the
oscillation is proportional to $\sqrt{a \, \kappa_b}$ while the decay
length, $\xi_t$, is independent of $a$, $\xi_t^{-1} = 1/( 4 \beta
\kappa_t) + (1 - \pi/4) / (\beta \kappa_b)$. It is therefore apparent,
that in the continuum limit, $a \to 0$, the oscillation period becomes
smaller and smaller, denoting a singular behaviour of the chain. This
is reminescent of the singular behaviour of rod-like chains which, in
the continuum limit, exhibit plectonemes on smaller and smaller
scales\cite{Moroz1,Mezard1}. From Fig. \ref{fig:phdiag}, it is
apparent that only if $\kappa_t$ is exactly zero, one remains in the
vanishingly small region of monotonic exponential decay when the
continuum limit is taken. In this case one recovers the WLC case with
persistence length $\xi_t \sim \beta \kappa_b$.

\begin{figure}[htbp]
\centerline{\includegraphics[width=2.4in]{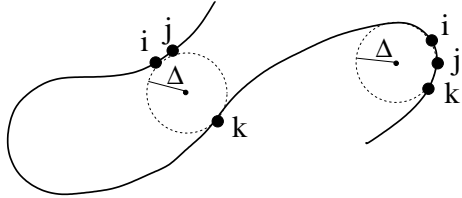}}
\caption{Sketch of a curve which is a viable centerline for a tube of
thickness $\Delta$. The radii of the circles through any triplet of
points, $r_{ijk}$ are not smaller than $\Delta$.}
\label{fig:0}
\end{figure}

We consider now the case of a polymer chain describing a thick
self-avoiding tube of uniform cross-section. The finite thickness
$\Delta$ of the polymer impacts on two distinct conformational
features.  First, it constrains the local radius of curvature to be
not less than $\Delta$ \cite{buck1,maddocks}. In addition, there is
also a non-local effect since any two portions of the tube, at a
finite arclength separation, cannot interpenetrate
\cite{buck1,maddocks}. In traditional beads-and-strings models it is
only this second effect that is taken into account through a pairwise
potential with a hard-core repulsion.  Interestingly, one needs to go
beyond pairwise interactions to account for the above mentioned
effects in discretized polymer chains \cite{Proteins,CPU,JSP,RMP}. In
fact, the requirement on the local radius of curvature can be enforced
by finding the radii of the circles going through any consecutive
triplet of points and ensuring that each of them is greater than
$\Delta$.  The non-local effect can be addressed within the same
framework by considering the minimum radius among circles going
through any non-consecutive triplet of points is also greater than
$\Delta$.  In summary the finite thickness, $\Delta$, of the
discretized tube requires that the radii of the circles going through
any triplet of distinct points have to be greater than $\Delta$ (see
Fig.  \ref{fig:0}) \cite{maddocks,JSP,RMP}.
In the present context, we are interested mainly in the local
thickness effects.  Therefore, we will consider the following reduced
Hamiltonian in the absence of torsional rigidity, involving only local
three-body constraints: ${\cal H}_1=\sum_i V(R_{i-1,i,i+1})$,
where $R_{ijk}$ is the radius of the circle going through the beads
$i$, $j$, $k$, and the potential $V(R)$ is $\infty$, if $R<\Delta$,
and $0$ otherwise. From simple geometric considerations, the local
tube constraint can be expressed in terms of the bond angle
distribution by imposing $\rho_{\theta}\left(\theta_i\right)=0$ if
$\theta_i > 2\arcsin\left(\frac{a}{2\Delta}\right)$. 
One can see that the thickness $\Delta$ plays a role similar to the
bending rigidity $\kappa_b$, in that they both induce the chain to be
locally more straight. Yet, the scaling behavior in the
$a\rightarrow0$ limit is different in the tube case, since the range
of $\theta$ angles most contributing to the average has now width
$\frac{a}{\Delta}$ (instead of $\sqrt{a/\beta \kappa_b}$), yielding
$\langle\cos\theta\rangle = 1 - {a^2}/{4\Delta^2}$ and
$\langle\sin\theta\rangle \sim {2a}/{3\Delta}$. Since $\Gamma =
\langle\cos\theta\rangle^2 \sim 1 - {a^2}/{2\Delta^2}$, the
tangent-tangent correlation function decays exponentially. This is
similar to the WLC case in the absence of torsional rigidity, but in
the tube model the correlation length scales in a different way $\xi_t
\sim \frac{\Delta^2}{a}$. This is a pathological behavior in the
continuous limit, since the correlation length diverges as
$a\rightarrow0$. The recipe by which the tube constraint is
implemented for a discrete chain ends up in preferentially sampling
straight continuous lines. It is natural to associate such ill
behavior to the degeneracy in the choice of the Frenet frame which
arises for a straight line conformation \cite{note}.  This can in fact
be cured by adding a torsional rigidity term to the tube constraint,
constraining the unphysical fluctuations of the binormal vectors
around straight line conformations. The Hamiltonian for this rod-like
thick polymer is:
\begin{equation}
{\cal H}_2 = \sum_{i}\ V(R_{i-1,i,i+1}) +
\frac{\kappa_t}{2a} \sum_i |\hat b_{i+1} - \hat b_{i}|^2\:.
\end{equation}
In this case, $\langle\cos\phi\rangle \sim 1-{a}/{2 \beta
\kappa_t}$ in the $a\rightarrow0$ limit, which implies $\xi_b = 2
\beta \kappa_t$ whereas we get $\Gamma \sim \frac{a^2}{36\beta^2
\kappa_t^2\Delta^2} \left[9\Delta^2-64\beta^2
\kappa_t^2\right]$. Thus, there are two different regimes, oscillatory
decay if $\kappa_t>\kappa_t^* \equiv \frac{3}{8}\frac{\Delta}{\beta}$,
and monotonic decay if $\kappa_t<\kappa_t^*$. In the latter case the
persistence length associated with the tangent-tangent correlation
function is controlled by {\em both} the tube thickness and the
torsional rigidity:
\begin{equation}
\xi_t = \frac{9\Delta^2}{16\beta \kappa_t}
\left[1+\sqrt{1-\left(\frac{8\beta \kappa_t}{3\Delta}\right)^2}\right] \: .
\label{correta}
\end{equation}
In the former case, the correlation length depends instead only on the
torsional rigidity, whereas the oscillation period depends also on the
tube thickness:
\begin{equation}
\xi_t = 4\beta \kappa_t \;\;, 
\chi = \frac{3\Delta}{2}
\left[1-\left(\frac{3\Delta}{8\beta \kappa_t}\right)^2\right]^{-1/2} \:. 
\label{corrxi}
\end{equation}
It can be seen from eq. (\ref{correta}), that by increasing the
torsional rigidity the tangent-tangent correlation length $\xi_t$,
initially controlled by the thickness $\Delta$ decreases until the
threshold $\kappa_t^*$ is reached. Above such threshold torsional
rigidity takes over and the tangent-tangent correlation length becomes
equal to twice the binormal-binormal correlation length, but the
thickness signature remains in the oscillatory behavior of the
tangent-tangent correlation function and in the period $\chi$.
\begin{figure}[htbp]
\centerline{\includegraphics[angle=0,width=3.5in,height=2.3in]{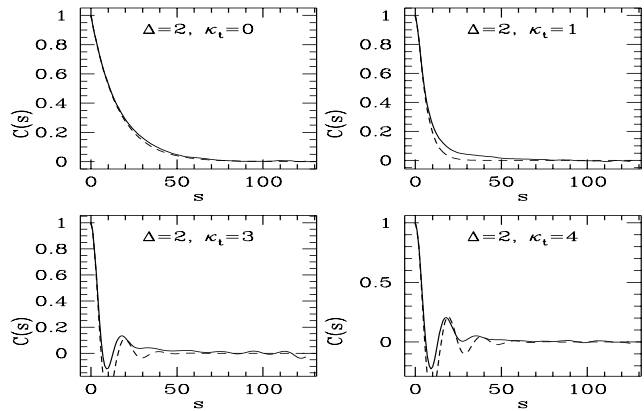}}
\caption{Tangent-tangent correlation, $C(s)$, as a function of the
arclength separation, $s$. The dashed and solid lines refer
respectively to the analytical results from eq. (\ref{tm}) and to the MC
simulations (maximum dispersion $\approx$ 0.01).}
\label{fig:delta2}
\end{figure}
The previous analysis, is in good semi-quantitative agreement with
data from MC simulations on the full model (i.e. where non-local
effects are retained), as visible in Fig. \ref{fig:delta2}. The
simulations were performed on chains of 128 equispaced beads through
the Metropolis acceptance of crankshaft and pivot moves. The
tangent-tangent correlations were measured by sampling structures at
intervals greater than the autocorrelation time.

To conclude, we have shown how the recently introduced thick-polymer
model can be regularized in the continuum limit by introducing a term
penalizing the geometrical torsion of the centerline. For any given
value of the polymer thickness there exists a torsional-rigidity
threshold separating the monotonic decay of the tangent-tangent
correlation from the oscillatory one. This highlights the rich
behaviour of thick-polymer models which combine features previuosly
observed in distinct polymer models, such as the worm- or rod-like
chains. The wide use of the latter in the context of single-molecule
experiments opens the possibility to use the physically-appealing
perspective of semi-flexible thick polymers to interpret the behaviour
of biopolymers.

We thank J. R. Banavar, T. X. Hoang and F. Seno for useful discussions
and acknowledge support from COFIN MURST 2003, FISR 2001, INFM and
EPSRC.


\begin{thebibliography}{99}
\bibitem{note} {\small A differentiable curve, $\vec{r}(s)$, $s$ being
the arclength parametrization, with thickness not smaller than
$\Delta$ satisfies: ${\rm sup}_{s_1,s_2} | \dot{\vec{r}}(s_1) -
\dot{\vec{r}}(s_2)| / |s_1 - s_2| \le \Delta^{-1}$ \cite{GMSM}. The
set of curves satisfying this bound has zero Wiener measure if the
latter includes only terms up to the second derivative in $\vec{r}(s)$
\cite{Freed}. Thus, higher-order derivatives are required for suitable
measures of tubes in the continuum limit.}
\end{thebibliography}
\end{document}